\documentclass[aps,prl,twocolumn,superscriptaddress,amsmath,floatfix,11pt,reprint]{revtex4-2} 
\usepackage[utf8]{inputenc}
\usepackage[T1]{fontenc}
\usepackage{graphicx}  
\usepackage{dcolumn}   
\usepackage{physics}
\usepackage{bm}        
\usepackage{amssymb}   
\usepackage{framed}
\usepackage{amsmath}
\usepackage{multirow}
\usepackage{hhline}
\usepackage{color}
\usepackage{xcolor}
\definecolor{bluegreen}{rgb}{0,0.2,0.8}
\usepackage{subfigure,amsmath,verbatim,moreverb}
\usepackage{tabularx}
\usepackage{adjustbox}
\usepackage{lipsum}
\usepackage{longtable}
\usepackage{booktabs}
\usepackage{adjustbox}
\usepackage{hyperref}
\UseRawInputEncoding
\usepackage{graphicx}
\usepackage{booktabs}
\usepackage{siunitx}
\usepackage{bbding}
\usepackage{amssymb}
\usepackage{pifont}
\usepackage[table]{xcolor}
\usepackage{ulem}  
\definecolor{lightcyan}{rgb}{0.88,1,1}

\usepackage{etoolbox}
\AtBeginEnvironment{align}{\setcounter{subeqn}{0}}
\newcounter{subeqn} %

\begin{document}
               
\title{
A Unified Dielectric-Dependent Hybrid Functional for Accurate Band Gaps across Dimensions
}

\author{Subrata Jana}
\email{subrata.niser@gmail.com}
\thanks{contributed equally to this work.}
\affiliation{Institute of Physics, Faculty of Physics, Astronomy and Informatics, Nicolaus Copernicus University in Toru\'n, ul. Grudzi\k{a}dzka 5, 87-100 Toru\'n, Poland}
 \affiliation{Institute of Advanced Studies, Nicolaus Copernicus University in Toru\'{n}, ul. Wile\'{n}ska 4, 87-100 Toru\'{n}, Poland} 

\author{Manoar Hossain}
\email{manoar.hossain@uni-paderborn.de} 
\thanks{contributed equally to this work.}
\affiliation{Paderborn Center for Parallel Computing, Paderborn University, Warburger Str. 100, 33098 Paderborn, Germany}

\author{Arghya Ghosh}
\affiliation{Department of Physics, Indian Institute of Technology, Hyderabad, India}

\author{Gabriel Chirchir}
\affiliation{Institute of Physics, Faculty of Physics, Astronomy and Informatics, Nicolaus Copernicus University in Toru\'n, ul. Grudzi\k{a}dzka 5, 87-100 Toru\'n, Poland}
 \affiliation{Institute of Advanced Studies, Nicolaus Copernicus University in Toru\'{n}, ul. Wile\'{n}ska 4, 87-100 Toru\'{n}, Poland} 

\author{Prasanjit Samal}
\affiliation{School of Physical Sciences, National Institute of Science Education and Research, An OCC of Homi Bhabha National Institute, Bhubaneswar 752050, India}

\author{Szymon \'Smiga}
\affiliation{Institute of Physics, Faculty of Physics, Astronomy and Informatics, Nicolaus Copernicus University in Toru\'n, ul. Grudzi\k{a}dzka 5, 87-100 Toru\'n, Poland}
 \affiliation{Institute of Advanced Studies, Nicolaus Copernicus University in Toru\'{n}, ul. Wile\'{n}ska 4, 87-100 Toru\'{n}, Poland} 

\date{\today}

\begin{abstract}

Predicting fundamental band gaps across material classes and dimensionalities remains a central challenge in electronic-structure theory. Here, we show that intrinsic dielectric screening provides a unified control parameter for nonlocal exchange from bulk to low-dimensional and heterogeneous materials. 
We introduce a geometry-independent dielectric response and incorporate it self-consistently into a nonempirical screened-dielectric-dependent hybrid functional. 
Benchmarks for 100 materials spanning bulk, two-dimensional, one-dimensional, and mixed-dimensional systems show near-$GW$ accuracy at the computational cost of generalized Kohn--Sham theory. 
These results reveal a screening--exchange--gap relation in which reduced dimensionality weakens intrinsic dielectric screening, strengthens nonlocal exchange, and drives the opening of fundamental gaps.
\end{abstract}

\maketitle

Accurate prediction of fundamental band gaps is essential for understanding and designing semiconductors, optoelectronic materials, and low-dimensional heterostructures~\cite{Curtarolo2013High,Jain2013Commentary}. However, despite sustained progress, a transferable description of band gaps across material classes and dimensionalities remains a major challenge in electronic-structure theory. Semilocal density-functional approximations (DFAs) systematically underestimate band gaps due to the lack of nonlocal exchange and the absence of the derivative discontinuity~\cite{Perdew1983Physical,Sham1983Density,Mori2008Localization}. In contrast, many-body $GW$ methods provide  reliable quasiparticle energies, but their high computational cost hinders applications to large supercells, surfaces, and reduced-dimensional systems~\cite{Onida2002Electronic,Wilhelm2021low,Graml2024low}. Hybrid functionals within generalized Kohn--Sham (gKS) theory offer a practical alternative~\cite{Perdew2017Understanding}, yet their transferability across dimensionalities remains limited~\cite{Wang2016Hybrid}. 

Dielectric-dependent (DD) hybrid functionals partially overcome this limitation by linking the long-range exact-exchange fraction to the electronic dielectric constant, enabling accurate band gaps for many bulk semiconductors and insulators~\cite{WeiGiaRigPas2018,Skone2016Nonempirical}. Their transferability to low-dimensional systems, however, remains limited~\cite{AshwinDahvydLeeor2019,Maria2023Transferable}. In reduced dimensions, the dielectric response becomes strongly anisotropic and spatially nonlocal~\cite{ZhengGovoniGalli2019}, and the computed macroscopic screening is further artificially diluted by vacuum in supercell calculations~\cite{Ghosh2025Advancing}. As a result, the screening depends on the chosen supercell rather than on the material itself, and defining screened exchange in a dimensionally consistent way remains a central challenge. 
    
In this Letter, we resolve this problem by introducing an intrinsic, vacuum-independent dielectric screening for systems treated within supercell approaches. The key idea is to separate the material contribution of polarization from geometric dilution caused by vacuum through an effective material volume extracted from the electron density. This yields a screening parameter that remains invariant with respect to the amount of vacuum included in the supercell and reduces to the usual macroscopic dielectric constant in the bulk limit. Incorporating this intrinsic screening into a dielectric-dependent range-separated hybrid functional leads to a unified screened-exchange framework applicable across dimensions and heterogeneous systems.

The performance of the resulting screened-exchange dielectric-dependent range-separated hybrid (SE-DD-RSH) functional is validated across a diverse set of three-dimensional (3D), two-dimensional (2D), one-dimensional (1D), and mixed-dimensional materials. Benchmarks show near-$GW$ accuracy, substantially improving over the Perdew--Burke--Ernzerhof (PBE) generalized gradient approximation (GGA)~\cite{perdewPRL96}, the Lebeda--Aschebrock--K\"{u}mmel (LAK) meta-GGA~\cite{Lebeda2024Balancing}, and the Heyd--Scuseria--Ernzerhof (HSE06) screened hybrid functional~\cite{heyd2003hybrid,heyd2004efficient}, 
while retaining the computational efficiency of gKS.

%

Typically, the quasiparticle (QP) band gap corrections in semiconductors and insulators are largely governed by nonlocal screened exchange. In the static limit of the $GW$ approximation, the self-energy reduces to the Coulomb-hole plus screened-exchange (COHSEX) form~\cite{Hybertsen1986Electron},
\begin{equation}
\Sigma^{\mathrm{COHSEX}}(\mathbf{r},\mathbf{r}')
=
\Sigma^{\mathrm{COH}}(\mathbf{r})\delta(\mathbf{r}-\mathbf{r}')+
\Sigma^{\mathrm{SEX}}(\mathbf{r},\mathbf{r}')
,
\end{equation}
where the Coulomb-hole (COH) contribution is predominantly local, whereas the nonlocal screened-exchange (SEX) term contains the statically screened Coulomb interaction $W(\mathbf{r},\mathbf{r}')$; this separation suggests a natural mapping within the gKS framework: nonlocal Fock exchange can approximate the SEX term, while semilocal exchange-correlation functionals account for the largely local COH contribution. In reciprocal space, the screened interaction in the long-wavelength limit satisfies~\cite{WeiGiaRigPas2018}
\begin{equation}
W(q\to0) = \frac{4\pi}{q^2}\frac{1}{\varepsilon_\infty},
\end{equation}
showing that macroscopic screening reduces the strength of long-range exchange by the factor $\varepsilon_\infty^{-1}$. This motivates hybrid functionals in which the fraction of exact exchange is linked to the dielectric response of the material.

To interpolate between the bare short-range interaction and the screened long-range limit, we employ a range-separated ansatz for the Coulomb kernel in reciprocal space~\cite{WeiGiaRigPas2018},
\begin{equation}
\begin{split}
\frac{4\pi}{\Omega G^{2}}
=
\frac{1}{\Omega}
\Bigl[
\frac{4\pi}{G^{2}}\{\alpha + (\gamma-\alpha)e^{-G^{2}/(4\mu^{2})}\}
\\
+
\frac{4\pi}{G^{2}}\{1-\alpha-(\gamma-\alpha)e^{-G^{2}/(4\mu^{2})}\}
\Bigr].
\label{eq2}
\end{split}
\end{equation}
This partitions the interaction into a nonlocal exact-exchange contribution and a complementary (semi)local part. The parameter $\mu$ defines the crossover wave vector between short- and long-range behavior. Physically, this form enforces the correct limiting behavior: in the limit $G\to\infty$, the exponential term vanishes, and the interaction approaches $\alpha\,4\pi/G^2$, recovering the bare Coulomb interaction at short distances. We therefore set $\alpha=1$~\cite{WeiGiaRigPas2018,Ghosh2025Advancing}, consistent with the exact condition $W(\mathbf{r},\mathbf{r}')\to v(\mathbf{r},\mathbf{r}')$ as $|\mathbf{r}-\mathbf{r}'|\to0$. In the opposite limit $G\to0$, the kernel reduces to $\gamma\,4\pi/G^2$, so that $\gamma$ controls the strength of long-range screened exchange and is thus identified with the inverse macroscopic dielectric constant $\varepsilon_{\infty}^{-1}$ within an effective scalar approximation to screening. 

For bulk solids, the electronic dielectric constant $\varepsilon_\infty$ is well defined in the $q\to0$ limit and can be reliably computed using density-functional perturbation theory (DFPT) with local-field effects~\cite{vaspoptics}. However, for low-dimensional systems modeled within supercells containing explicit vacuum, the calculated macroscopic dielectric constant $\varepsilon_\infty^{\mathrm{cell}}$ acquires an artificial dependence on the supercell size and gradually approaches unity as the vacuum thickness increases~\cite{Ghosh2025Advancing}. This artifact originates from the volume averaging of the macroscopic polarization, which includes vacuum regions that do not contribute to the screening. To recover an intrinsic, geometry-independent dielectric response, we separate the material contribution to polarization from this geometric dilution. The macroscopic polarization is given by
\begin{equation}
P = \frac{1}{{\Omega_{\mathrm{cell}}}}\int_{{\Omega_{\mathrm{cell}}}} p(\mathbf{r})\, d^3 r,
\end{equation}
where $p(\mathbf{r})$ is the microscopic polarization density. In a supercell geometry, $p(\mathbf{r})$ is finite only within the material region and vanishes in vacuum where the electronic susceptibility is zero. Denoting the effective material volume by $\Omega_{\mathrm{eff}}$ and the total supercell volume by $\Omega_{\mathrm{cell}}$, one finds
\begin{equation}
P^{\mathrm{cell}}
=
\frac{1}{\Omega_{\mathrm{cell}}}
\int_{\Omega_{\mathrm{eff}}} p(\mathbf{r})\, d^3 r
=
\frac{\Omega_{\mathrm{eff}}}{\Omega_{\mathrm{cell}}}
P^{\mathrm{mat}} ,
\end{equation}
where $P^{\mathrm{mat}}$ is the intrinsic polarization averaged over the material region. The macroscopic polarization extracted from the supercell is thus reduced by a purely geometric dilution factor. Since the dielectric susceptibility is defined by $P = \chi E$ in the linear-response regime, the same factor reduces the macroscopic susceptibility and hence the dielectric constant. This leads to 
\begin{equation}
\varepsilon_\infty^{\mathrm{eff}}
=
1
+
\eta(\varepsilon_\infty^{\mathrm{cell}} - 1),
\qquad
\eta = \frac{\Omega_{\mathrm{cell}}}{\Omega_{\mathrm{eff}}},
\label{Eq:effective}
\end{equation}
where $\varepsilon_\infty^{\mathrm{eff}}$ represents the intrinsic effective electronic screening of the material, independent of vacuum size. 

The effective material volume is determined from the electronic density through
\begin{equation}
\Omega_{\mathrm{eff}} = 
\int_{\Omega_{\mathrm{cell}}} w(\mathbf{r})\, d^3 r,
\qquad
w(\mathbf{r}) = \operatorname{erf}\!\left[\frac{n(\mathbf{r})}{n_c}\right],
\end{equation}
where the weighting function $w(\mathbf{r})$ distinguishes material regions from vacuum based on the electronic density $n(\mathbf{r})$ and cutoff density $n_c = 6.96 \times 10^{-4}\,e/\mathrm{bohr}^3$~\cite{Singh2025Simplified}. By construction, this definition ensures: (i) the correct bulk limit $\eta = 1$; (ii) $\varepsilon_\infty^{\mathrm{eff}} = 1$ in the pure vacuum limit; and (iii) invariance of $\varepsilon_\infty^{\mathrm{eff}}$ with respect to supercell size. For layered 2D systems, identifying $\eta$ with the ratio of supercell height to effective layer thickness recovers the standard capacitor model~\cite{Laturia2018dielectric,Ghosh2025Advancing} for two-dimensional materials.

In low-dimensional systems, the dielectric response is intrinsically anisotropic. To define an effective scalar screening parameter, we retain only the physically relevant components of the dielectric tensor and replace vacuum-dominated directions by unity,
\begin{equation}
\tilde{\boldsymbol{\epsilon}}=
\begin{cases}
\boldsymbol{\epsilon} & (3\mathrm{D}),\\
\operatorname{diag}(\epsilon_{xx},\epsilon_{yy},1) & (2\mathrm{D}),\\
\operatorname{diag}(1,1,\epsilon_{zz}) & (1\mathrm{D}),
\end{cases}
\end{equation}
and define $\varepsilon_\infty^{\mathrm{cell}} =\frac{1}{3}\operatorname{Tr} \tilde{\boldsymbol{\epsilon}}$.
This framework captures the physically relevant screening channels while removing spurious vacuum contributions.

The range-separation parameter $\mu$ defines a crossover length $r_c \sim \mu^{-1}$ between bare and screened exchange. Since electronic screening is governed by the characteristic interelectronic spacing, this crossover should scale with the average electronic density. We therefore relate $\mu$ to the density-averaged Seitz radius
\begin{equation}
\langle r_s \rangle =
\frac{\int_{\Omega_{\mathrm{cell}}} w(\mathbf{r}) r_s(\mathbf{r})\, d^3 r}
{\int_{\Omega_{\mathrm{cell}}} w(\mathbf{r})\, d^3 r},
\qquad
r_s(\mathbf{r}) = \left(\frac{3}{4\pi n(\mathbf{r})}\right)^{1/3},
\end{equation}
using
\begin{equation}
\mu =
\frac{a_1}{\langle r_s \rangle}
+
\frac{a_2 \langle r_s \rangle}
{1 + a_3 \langle r_s \rangle^2}
\label{mu-eff}
\end{equation}
with $a_1 = 1.91718$, $a_2 = -0.02817$, and $a_3 = 0.14954$~\cite{Jana2023simple}.
This reflects the expected trend that denser systems (smaller $\langle r_s \rangle$) possess shorter intrinsic screening lengths and therefore larger $\mu$. The same weighting function suppresses vacuum contributions, ensuring that both $\gamma$ and $\mu$ are intrinsic and independent of the supercell geometry. The resulting functional is referred to as the SE-DD-RSH. We note that the functional form of Eq.~\eqref{mu-eff} has been successfully employed in earlier studies of solids, including surfaces and interfaces, providing a robust and transferable prescription for $\mu$~\cite{Jana2023simple,Singh2025Simplified,Ghosh2026Accurate}.

Since $\varepsilon_\infty^{\mathrm{eff}}$ depends on the electronic structure, the dielectric-dependent parameters are determined self-consistently: starting from an initial guess for $\varepsilon_\infty^{\mathrm{eff}}$, we construct $\gamma$ and $\mu$, compute $\varepsilon_\infty^{\mathrm{cell}}$ including local-field effects, update $\varepsilon_\infty^{\mathrm{eff}}$ through the corresponding rescaling relation, and iterate to convergence (see Fig.~S1 of the Supplemental Material~\cite{support}). The proposed SE-DD-RSH functional thereby provides a unified and nonempirical description of electronic 
screening across bulk, surface, and low-dimensional systems within a single gKS framework.

%

To assess the performance of SE-DD-RSH for band-gap prediction across dimensionalities and material classes, we consider a diverse benchmark set comprising bulk and quasi-two-dimensional solids, monolayers, one-dimensional nanostructures, surfaces, and mixed-dimensional hybrid systems. All density-functional calculations (PBE, LAK, HSE06, and SE-DD-RSH) are performed using VASP~\cite{vasp1,vasp2,vasp3,vasp4}. Quasiparticle $GW$ calculations for three- and two-dimensional systems are carried out in VASP, while one-dimensional systems are treated using BerkeleyGW~\cite{BerkeleyGW} together with Quantum ESPRESSO~\cite{Giannozzi_2017}. The dielectric rescaling factor $\eta$ and range-separation parameter $\mu$ are obtained from PBE charge densities~\cite{perdewPRL96}. Further details and the workflow (Fig.~S1) are given in the Supplemental Material (SM)~\cite{support}.

The robustness of the intrinsic screening parameters with respect to the supercell vacuum is examined using an isolated $h$-BN monolayer. 
Increasing the out-of-plane lattice spacing strongly alters the supercell dielectric constant $(\varepsilon_{\mathrm{cell}})$ and the rescaling factor $\eta$. In contrast, the intrinsic screening parameters, $\gamma~(\varepsilon_{\mathrm{eff}}^{-1})$ and $\mu$, remain nearly unchanged (Table~S1 and Fig.~S2 of SM~\cite{support}). 
This confirms that the density-based rescaling effectively removes the artificial vacuum dependence of the dielectric response, thereby yielding stable intrinsic screening parameters for reduced-dimensional systems. The resulting values 
of $\eta$, $\gamma$, and $\mu$ for all benchmark systems are listed in the Supplemental Material~\cite{support} (Tables~S2--S5).


\begin{table*}[t]
\caption{Statistical errors (in eV for ME and MAE, and \% for MARE) of different functionals with respect to $GW$ band gaps across 3D (33 materials), 2D (33 materials), 1D (34 materials), and the full dataset (100 materials). The best values are in  bold.}
\label{tab-bg2}
\begin{ruledtabular}
\begin{tabular}{lccc|ccc|ccc|ccc}
 & \multicolumn{3}{c}{3D} & \multicolumn{3}{c}{2D} & \multicolumn{3}{c}{1D} & \multicolumn{3}{c}{All} \\
\cline{2-13}
Method
& ME & MAE & MARE
& ME & MAE & MARE
& ME & MAE & MARE
& ME & MAE & MARE \\
\hline
PBE        & -2.37 &  2.37 & 50.80 & -1.31 &  1.31 & 48.39 & -2.81 &  2.81 & 61.96 & -2.17 &  2.17 & 53.80 \\
LAK        & -1.71 &  1.73 & 31.45 & -0.95 &  0.95 & 34.05 & -2.47 &  2.47 & 55.06 & -1.72 &  1.72 & 40.33 \\
HSE06      & -1.11 &  1.13 & 16.23 & -0.67 &  0.67 & 22.88 & -2.03 &  2.03 & 44.73 & -1.27 &  1.28 & 28.11 \\[0.2 cm]
SE-DD-RSH  &  0.13 &  {\bf 0.30} &  {\bf 7.39} &  0.19 &  0.35 & 12.06 &  0.61 &  0.84 & {\bf 17.84} &  0.31 &  0.50 & {\bf 12.48} \\
SE-DD-RSH0 &  {\bf 0.12} &  0.47 & 15.46 & {\bf -0.01} &  {\bf 0.20} &  {\bf 7.74} &  {\bf 0.20} &  {\bf 0.69} & 17.95 &  {\bf 0.11} &  {\bf 0.46} & 13.76 \\
\end{tabular}
\end{ruledtabular}
\end{table*}
%

\begin{figure}
\centering
\includegraphics[width=8.5cm, height=5.8cm]{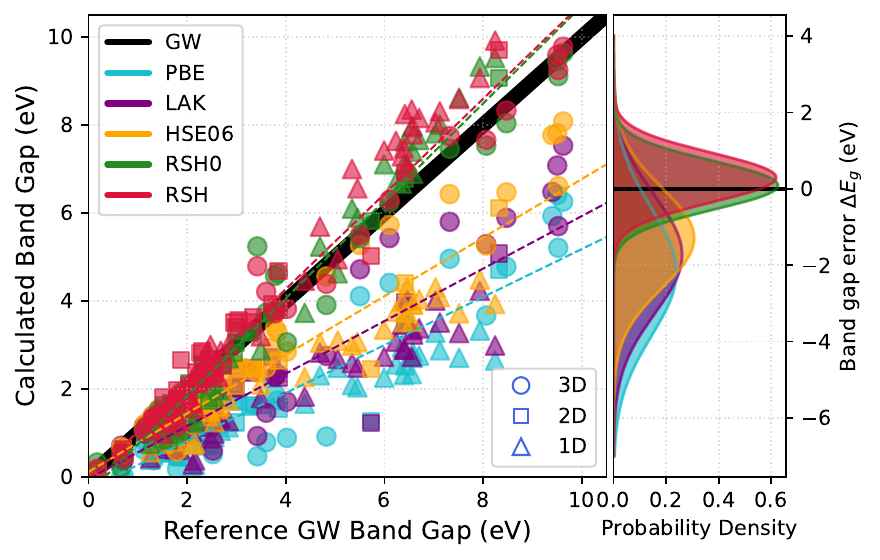}
\caption{Comparison of calculated band gaps with reference $GW$ values and corresponding error distributions. Left: Calculated band gaps from the considered functionals plotted against quasiparticle $GW$ band gaps. The dashed line denotes perfect agreement. Right: Gaussian fits to the error distributions relative to $GW$.} 
\label{fig:band-gap}
\end{figure}

We benchmark SE-DD-RSH against a representative hierarchy of semilocal and hybrid functionals, using $GW$ quasiparticle band gaps as reference. We also consider a computationally simpler single-shot variant, SE-DD-RSH0, in which the dielectric screening is evaluated from the PBE and kept fixed; the band gap is then computed without self-consistent updating of the screening, using PBE wavefunctions and charge density~\cite{support} (see Fig.~S1 of Ref.~\cite{support}). Across the full dataset, both SE-DD-RSH and SE-DD-RSH0 substantially reduce the systematic underestimation of PBE, LAK, and HSE06, yielding significantly smaller mean errors (ME) and mean absolute errors (MAE) across 3D, 2D, and 1D materials (Table~\ref{tab-bg2}). 

\begin{figure}
\centering
\includegraphics[width=8.5cm, height=7cm]{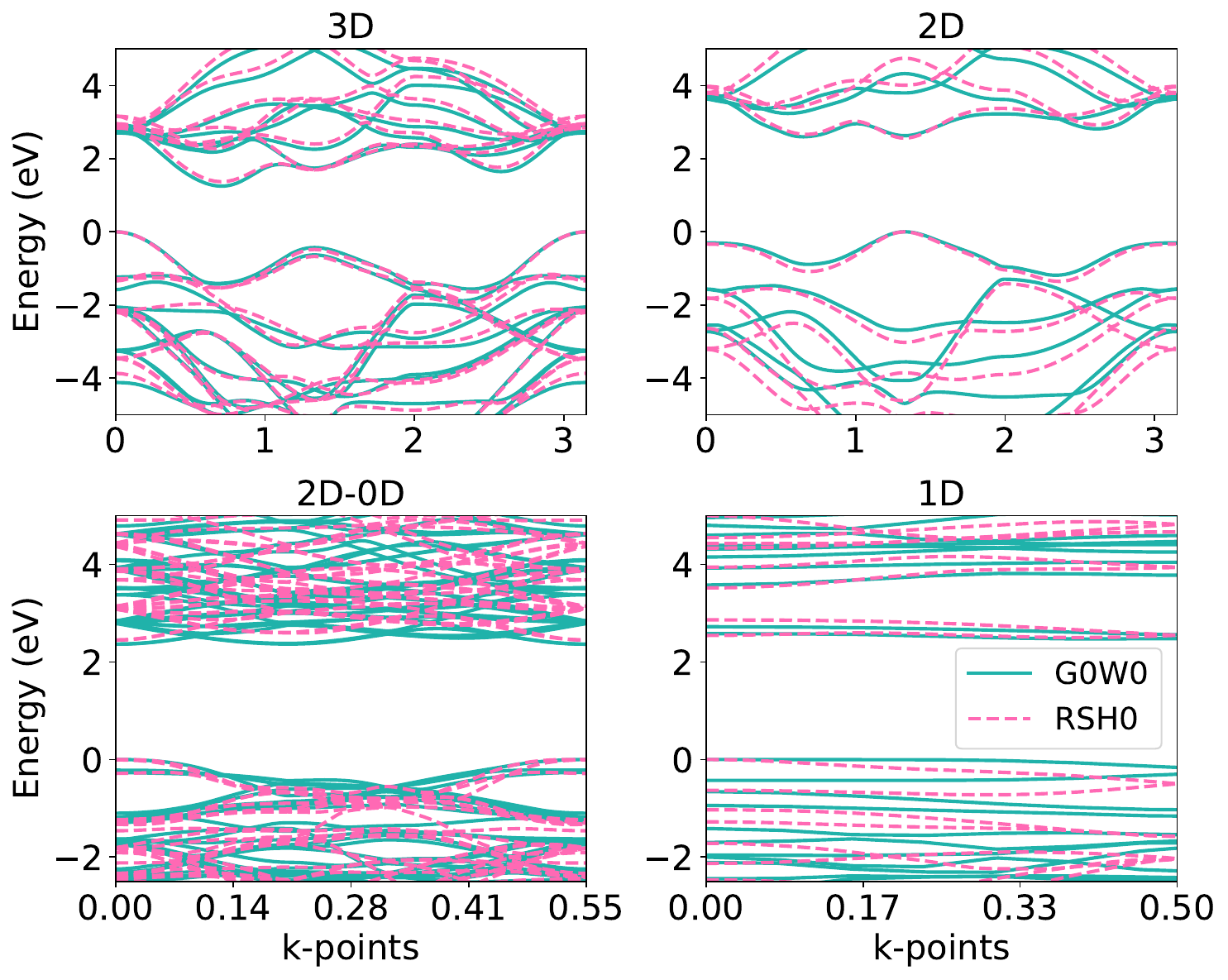}
\caption{SE-DD-RSH0 and QP band structures for 3D, 2D, mixed-dimensional (2D-0D) material, and 1D structure. QP band structure data for the 2D-0D material is taken from Ref.~\cite{Schebek2025Efficient}. All results are for MoS$_2$ and pyridine@MoS$_2$.}
\label{fig:band-gap-mixed}
\end{figure}

For bulk and quasi-two-dimensional systems, where the effective volume coincides with the simulation-cell volume ($\eta = 1$), SE-DD-RSH yields band gaps in close agreement with $GW$ over a broad range of materials, including wide-gap insulators, semiconductors, perovskites, antiferromagnets, and layered compounds (Table~S2~\cite{support}). While LAK partially reduces the large underestimation of PBE, it remains qualitatively inaccurate for insulators and oxides, and HSE06 performs reasonably well, mainly for small- and medium-gap materials. In contrast, SE-DD-RSH achieves a MAE of $\sim 0.3$~eV, compared with $1.7$~eV for LAK and $1.1$~eV for HSE06 (Table~\ref{tab-bg2}). This indicates that dielectric-dependent screened exchange provides a more reliable description of the quasiparticle gap correction in extended solids than fixed-fraction hybrid functionals.

The improvement becomes more pronounced in reduced dimensions, where the dielectric screening is weaker, anisotropic, and strongly influenced by confinement. For monolayers, heterostructures, and surfaces, including transition-metal dichalcogenides, $h$-BN, phosphorene, Janus materials, and magnetic systems such as CrBr$_3$, both SE-DD-RSH and SE-DD-RSH0 consistently outperform HSE06 and LAK (Table~S3 of Ref.~\cite{support}). In particular, the single-shot variant performs remarkably well in two dimensions, yielding a MAE of $\sim 0.2$~eV and a MARE of 7.7\% across the 2D dataset (Table~\ref{tab-bg2}). This shows that the dominant improvement originates from the intrinsic screening construction itself rather than from full self-consistency.

Remarkably, a similar trend is observed for one-dimensional systems such as nanoribbons, nanotubes, and nanowires (Table~S4 of Ref.~\cite{support}). In these systems, PBE, LAK, and HSE06 systematically underestimate the confinement-induced band-gap enhancement arising from reduced dielectric screening, with errors increasing as confinement strengthens. By contrast, SE-DD-RSH and SE-DD-RSH0 accurately recover this confinement-driven gap opening and substantially reduce the relative error with respect to $G_0W_0$. For these materials, SE-DD-RSH0 yields the smallest MAE ($0.7$~eV), while SE-DD-RSH gives the lowest MARE (17.8\%) (Table~\ref{tab-bg2}). These results demonstrate that fixed-parameter functionals fail to account for the strengthening of screened exchange in low dimensions, whereas the present framework naturally adapts to screening environments.

Overall, the deviations from $GW$ remain consistently small, indicating that SE-DD-RSH(0) captures the essential screening physics of the quasiparticle self-energy within a gKS framework. As summarized in Table~\ref{tab-bg2}, across a dataset of 100 materials spanning 3D, 2D, and 1D systems, SE-DD-RSH(0) achieves an overall MAE of $\sim 0.5$~eV (MARE $\sim 12\%$), representing nearly a threefold reduction relative to HSE06.

Figure~\ref{fig:band-gap} summarizes these trends for the complete benchmark set. PBE, LAK, and HSE06 exhibit large deviations from the diagonal and broad error distributions centered at negative values, reflecting systematic underestimation of band-gaps. In contrast, SE-DD-RSH and SE-DD-RSH0 lie very close to the diagonal and show narrow error distributions centered near zero. This confirms that incorporating nonempirical, material-dependent screening leads to more accurate and robust band-gap predictions across diverse materials and dimensions. 


The excellent computational efficiency and accuracy of SE-DD-RSH0 are particularly suitable for heterogeneous and mixed-dimensional systems, where fully self-consistent hybrid-functional calculations are computationally demanding. As a proof of concept, we consider pyridine@MoS$_2$, for which SE-DD-RSH0 yields a band gap (Table~S5~\cite{support}) in close agreement with the corresponding QP reference~\cite{Schebek2025Efficient,Gonzalez2024Hybrid}. Remarkably, as shown in Fig.~\ref{fig:band-gap-mixed}, SE-DD-RSH0 also reproduces representative QP band structures across bulk, two-dimensional, mixed-dimensional, and one-dimensional systems, indicating that the present framework captures the interplay of dielectric screening across chemically and spatially distinct environments.

In conclusion, despite important progress in hybrid-functional development, a broadly transferable, nonempirical framework capable of delivering near-$GW$ band gaps across bulk, low-dimensional, and mixed-dimensional materials has remained an open challenge.
Here, we introduce a simple, nonempirical, and easily implementable hybrid functional that provides a unified and transferable framework for predictive electronic-structure simulations of bulk, low-dimensional, and nanoscale heterogeneous materials. 
By explicitly incorporating material-specific dielectric screening, the proposed approach yields band gaps in quantitative agreement with $GW$ calculations, without empirical tuning and at a fraction of the computational cost. 
The functional can be interpreted as a static, nonempirical analogue of the $GW$ self-energy in the COHSEX limit, retaining the essential physics of screened exchange while avoiding the unfavorable computational scaling of many-body perturbation theory. 
To the best of our knowledge, this is the first unified hybrid-functional framework capable of achieving near-$GW$ accuracy for band gaps across diverse materials and dimensionalities. The present formulation thus enables efficient and reliable large-scale simulations of van der Waals heterostructures, surfaces, and quantum-confined systems, opening new directions for predictive electronic-structure studies.

All data supporting the findings of this study are available within the paper and Supplemental Material. Structural input files are available at [link to be provided upon publication]. Additional data are available from the corresponding author upon reasonable request. M.H. acknowledges computing time provided on the high-performance computers Noctua 2 and Otus~\cite{manoar2025otus} at the NHR Center PC$2$. The NHR center is jointly supported by the Federal Ministry of Education and Research and the state governments participating in the NHR (www.nhr-verein.de). S.~J, G.~C. and S.~S. acknowledge financial support from the National Science Center, Poland (Grant No. 2021/42/E/ST4/00096). S.J. gratefully acknowledges Fabien Tran for discussions on some aspects of the implementation.

\twocolumngrid

\bibliography{reference}
\end{document}